\documentclass[prl,twocolumn,notitlepage]{revtex4-2}
\usepackage[english]{babel}
\usepackage{amsfonts}
\usepackage{physics}
\usepackage{graphicx}
\usepackage{placeins}
\usepackage{url}
\usepackage{caption}
\usepackage{graphicx}
\usepackage{dcolumn}
\usepackage{bm}

\usepackage{mathtools}
\usepackage{color}
\begin{document}


\title{On the robustness of the \textit{in vivo} cyanobacterial circadian clock}

\author{Dorota Youmbi Fouego${}^{\natural, \diamond} $}

\author{ Sophie de Buyl ${}^{\diamond, \flat}$}
\email[Correspondence to: ]{sdebuyl@vub.be}
\affiliation{${}^\natural$ Laboratory of Modelling and Simulation in Engineering, Biomimetics and Prototypes and TWAS Research Unit, Department of Physics, Faculty of Science, University of Yaound\'e I, Po. Box 812, Yaound\'e, Cameroon}
\affiliation{${}^{\diamond} $ Applied Physics Research Group, Physics Department, Vrije Universiteit Brussel, Brussels 1050, Belgium}
\affiliation{${}^\flat$ Interuniversity Institute of Bioinformatics in Brussels, Brussels 1050, Belgium}





\date{\today}

\begin{abstract}
 We propose a revisited version of the \textit{in vivo} model of the cyanobacterial circadian clock. Our aim is to address the lack of robustness predicted for the mutant cyanobacteria without transcriptional regulation of the original model. For this, we rely on an \textit{in vitro} model of the clock describing explicitly the hexameric structure of the core protein of the clock. Our model is able to reproduce oscillatory behavior for the mutant, as observed experimentally, without finely tuned parameters. 
\end{abstract}

\maketitle


\section{\label{sec:intro}Introduction}

Circadian clocks are endogenous oscillators allowing organisms to synchronize their physiological activities and behavior with the time of the day \cite{goldbeter_biochemical_1996,goldbeter_systems_2012}. They play a role in most living organisms from bacteria to humans. Cyanobacteria possess one of the simplest known circadian clock and serve as one of the model organisms to study mechanisms leading to endogenous oscillations of about 24 hours. One particularly beautiful property of the cyanobacterial circadian clock is that it can be reconstituted in a test tube. If one mixes adenosine triphosphate (ATP) with the three key proteins of the clock, namely KaiC, KaiB and KaiA, autonomous oscillations with a period of about 24 hours are observed during several days \cite{nakajima_reconstitution_2005, tomita_no_2005,rust_ordered_2007,ito_autonomous_2007}

The mechanism of the clock is essentially an ensemble of phosphorylation/dephosphorylation reactions of KaiC which is the core protein of the clock. The phosphorylation state of KaiC contains the information about the phase of the clock. More specifically, the structure of KaiC is a homo-hexamer in the shape of a double-doughnut which can be phosphorylated on two residus on each monomer. The role of KaiA is to promote phosphorylation while KaiB, when bound to KaiC, inhibits phosphorylation by sequestring KaiA. Those phosphorylation/dephosphorylation reactions can be reproduced in the test tube. An additional mechanism comes at play \textit{in vivo} as KaiC autoregulates its own production via a negative feedback on its own mRNA production. There are therefore two oscillatory processes underlying the functioning of the clock, the post-translational regulation (PTR) which is observed \textit{in vitro}, and the transcriptional translational regulation (TTR) consisting of the negative feedback on mRNA production. 

A mathematical model of the \textit{in vitro} clock relying on careful experiments measuring all kinetic rates and concentrations of proteins involved was proposed in \cite{rust_ordered_2007}. This simple model beautifully reproduces oscillations with a period of about 24 hours, without any parameter space exploration as all parameters have been measured. 

In \cite{teng_robust_2013}, we proposed an extension to the \textit{in vivo} case. This model showed that PTR regulation is sufficient to generate oscillations, as observed in the experiments with mutant cyanobacteria lacking the TTR regulation. It also showed that the transcriptional regulation helps maintaining synchrony in a population of growing cyanobacteria. However, the model requires finely tuned parameters to lead to oscillatory behavior of the mutant cyanobacteria lacking the transcriptional regulation.  
The \textit{in vitro} model it is based on did not explicitly described the hexameric structure of the KaiC protein. 
More recently, a more realistic model of the \textit{in vitro} clock has been proposed in \cite{lin_mixtures_2014}. It describes explicitly the hexameric nature of KaiC and the binding of KaiB to KaiC. 
We propose here to build an \textit{in vivo} model based on this more realistic \textit{in vitro} model. Our model is able to produce robust oscillations for the wild type cyanobacteria as well as for the mutants lacking transcriptional regulation. 

The structure of the work is as follows. We first briefly introduce the \textit{in vitro} model of \cite{lin_mixtures_2014} for completeness and to introduce notations. Then we describe our \textit{in vivo} model of the cyanobacterial circadian clock. We analyze the robustness of our model with respect to parameter variations, and compare it with the model proposed in \cite{teng_robust_2013}. We end with a short discussion on the role of transcriptional regulation. 

\section{The \textit{in vitro} hexameric model \label{sec:model} }

The hexameric model of the clock describes the interconversions of the different phosphoforms of KaiC as well as the binding and unbinding of KaiC to KaiB. When KaiC is bound to KaiB, it forms the KaiB.KaiC complex which can undergo to same (de)phosphorylation reactions as KaiC. The nonlinearity in the system comes from the fact that the reaction rates depend on the state of the system. When KaiA is bound to KaiC, it enhances the auto-kinase activity of KaiC. Although KaiA is not described explicitly, it impacts the (de)phosphorylation rates as KaiC binds differentially to KaiA in its different phosphoforms. Similarly, KaiB is implicitly taken into account and antagonizes the effect of KaiA in a KaiC phosphoform dependent manner. Essentially, when KaiC subunits are in the $S$ form, KaiB will bind to KaiC and sequestrate KaiA, thereby repressing phosphorylation. More details are provided in the appendix. A scheme of the model is depicted in Fig.~\ref{fig:figure1}.
\begin{figure}[h]
\includegraphics[width = .4\textwidth]{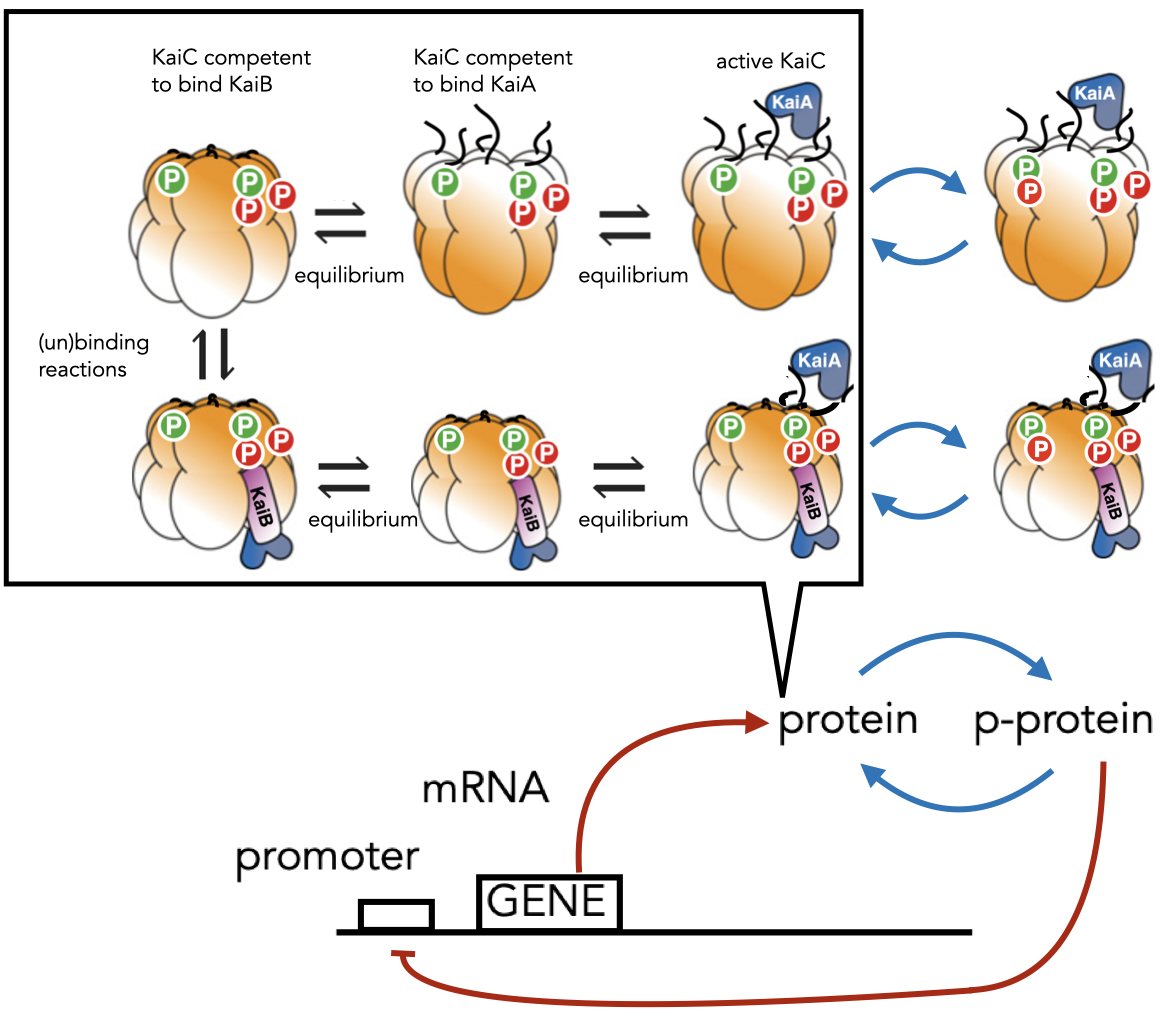}
\caption{ \label{fig:figure1} Schematic representation of the model, adapted from \cite{lin_mixtures_2014}. The representation of the configuration of the active form of KaiB.KaiC is speculative. }
\end{figure}
Each subunit of KaiC undergoes independent autokinase and autophosphatate reactions at two sites, serine 431 (S) and threonine 432 (T) which can be phosphorylated or not. Therefore, each subunit can be in four forms, namely: the phosphorylated form at S, the phosphorylated form at T, the doubly phosphorylated form at S and T (denoted D), and the U form which is the non-phosphorylated form. The different phosphoforms are denoted by KaiC${}_{ijk}$ where 
$i$ represents the number of subunits phosphorylated at T only, $j$ represents the number of phosphorylated subunits at S only and $k$ represents the number of doubly phosphorylated subunits.  By construction, the variables $KaiC_{ijk}$ which are physical have indices obeying $0< i, j, k< n$ and the number of unphosphoryated subunits is given by $n-(i+j+k)$. For the hexamer model $n=6$, but to study theoretically the effect of changing the number of KaiC subunits, the model was constructed with an arbitrary number subunits denoted by $\mathnormal{n}$. For instance, the unphosphorylated KaiC is $KaiC_{000}$ and the fully doubly phosphorylated form is $KaiC_{00n}$. One should note that the dynamical variables constructed by assuming that the different subunits of the hexamer are indistinguishable. 
KaiC can undergo potentially (de)phosphorylation reactions which will increase or decrease the values of $i, j, k$ by one unit. Additionally, $KaiC_{i,j,k}$ can bind to KaiB, forming the complex $KaiB.KaiC_{i,j,k}$ which undergoes the same (de)phosphorylation reactions. The model is given by Eqs. (\ref{Eq.invitro}). The dependence of the reaction rates on the KaiC phosphoforms are given in the appendix, see Eqs.~\ref{phosphorate} and \ref{bindrate}.


\begin{widetext}
\begin{eqnarray} \label{Eq.invitro}
\left.\ \frac{ d \text{KaiC}_{i,j,k}}{dt} \right|^{in \, vitro} &=& 
k_{UT}^{i-1,j,k}(\mathnormal{n} +1-(i+j+k)) \text{KaiC}_{i-1,j,k} + 
k_{US}^{i,j-1,k}(\mathnormal{n} +1-(i+j+k))\text{KaiC}_{i,j-1,k}\\
&+& k_{TD}^{i+1,j,k-1}(i+1) \text{KaiC}_{i+1,j,k-1}+
k_{TU}^{i+1,j,k} (i+1) \text{KaiC}_{i+1,j,k}  \nonumber  \\
&+& k_{DT}^{i-1,j,k+1} (k+1) \text{KaiC}_{i-1,j,k+1}+
k_{DS}^{i,j-1,k+1}(k+1) \text{KaiC}_{i,j-1,k+1}  \nonumber  \\
&+& k_{SD}^{i,j+1,k-1}(j+1) \text{KaiC}_{i,j+1,k-1} + 
k_{SU}^{i,j+1,k}(j+1) \text{KaiC}_{i,j+1,k}  \nonumber  \\
&-& ((n-i-j-k) \, ( k_{UT}^{i,j,k} + k_{US}^{i,j,k} )+ i \, k_{TD}^{i,j,k} + i \, k_{TU}^{i,j,k} + k \, k_{DT}^{i,j,k} + k \, k_{DS}^{i,j,k} + j \, k_{SD}^{i,j,k} + j \, k_{SU}^{i,j,k} ) \text{KaiC}_{i,j,k}  \nonumber \\
&-& k_{\text{on,} B} F^{i,j,k}_B \,  \text{KaiC}_{i,j,k}  \nonumber  \\ 
&+& \, k_{\text{off},B} \, \text{KaiB.KaiC}_{i,j,k} \nonumber  \\
\left.\  \frac{d \text{KaiB.KaiC}_{i,j,k} }{dt}  \right|^{in \, vitro}  &=&  k_{UT}^{i-1,j,k}(\mathnormal{n} +1-(i+j+k)) \text{KaiB.KaiC}_{i-1,j,k} + 
k_{US}^{i,j-1,k}(\mathnormal{n} +1-(i+j+k))\text{KaiB.KaiC}_{i,j-1,k} \nonumber \\
&+& k_{TD}^{i+1,j,k-1}(i+1) \text{KaiB.KaiC}_{i+1,j,k-1}+
k_{TU}^{i+1,j,k} (i+1) \text{KaiB.KaiC}_{i+1,j,k}  \nonumber  \\
&+& k_{DT}^{i-1,j,k+1} (k+1) \text{KaiB.KaiC}_{i-1,j,k+1}+
k_{DS}^{i,j-1,k+1}(k+1) \text{KaiB.KaiC}_{i,j-1,k+1}  \nonumber  \\
&+& k_{SD}^{i,j+1,k-1}(j+1) \text{KaiB.KaiC}_{i,j+1,k-1} + 
k_{SU}^{i,j+1,k}(j+1) \text{KaiB.KaiC}_{i,j+1,k}  \nonumber  \\
- ((n-&i&-j-k) \, ( k_{UT}^{i,j,k} + k_{US}^{i,j,k} )+ i \, k_{TD}^{i,j,k} + i \, k_{TU}^{i,j,k} + k \, k_{DT}^{i,j,k} + k \, k_{DS}^{i,j,k} + j \, k_{SD}^{i,j,k} + j \, k_{SU}^{i,j,k} ) \, \text{KaiB.KaiC}_{i,j,k} \nonumber \\
&-& k_{\text{off,} B} \, \text{KaiB.KaiC}_{i,j,k} \nonumber  \\ 
&+& \, k_{\text{on,B} } F^{i,j,k}_B \, \text{KaiC}_{i,j,k} \nonumber 
\end{eqnarray}
\end{widetext}

\section{The \textit{in vivo} hexameric model \label{sec:model} }

To construct an \textit{in vivo} model, we rely on the \textit{in vitro} model of \cite{lin_mixtures_2014} described in the section above and add terms for production, degradation and dilution similarly to what is proposed in \cite{teng_robust_2013}. Our model is given by the Eqs. \ref{Eq.vivo}. 
\begin{widetext}
\begin{eqnarray} 
\frac{d \text{KaiC}_{ijk} }{dt}  &=& \left.\  \frac{ d \text{KaiC}_{ijk}}{dt} \right|^{in \, vitro}  - V_d \text{KaiC}_{ijk} - V \frac{\text{KaiC}_{ijk}}{K + \text{KaiC}_{ijk}}\, \, \, \text{for } i,j,k \neq 0,0,0 \label{Eq.vivo} \\
\frac{d \text{KaiB.KaiC}_{ijk} }{dt}  &=& \left.\  \frac{ d \text{KaiB.KaiC}_{ijk}}{dt} \right|^{in \, vitro}  \, \, \text{for } i,j,k \neq 0,0,0 \nonumber \\
\frac{ d \text{KaiC}_{000}}{dt}   &=& \left.\  \frac{ d \text{KaiC}_{000}}{dt} \right|^{in \, vitro} + K_s \text{mRNA}  - V_d \text{KaiC}_{000} - V \frac{\text{KaiC}_{000}}{K + \text{KaiC}_{000}} \nonumber \\
 \frac{ d \text{mRNA} }{dt}  &=&  V_s \frac{K_i^4}{K_i^4 + (\sum_{j \neq 0} \text{KaiB.KaiC}_{ijk})^4} -  V_m \frac{\text{mRNA}}{K_m +\text{mRNA}}  \nonumber
\end{eqnarray}
\end{widetext}
Firstly, we consider explicitly KaiC mRNA as a dynamical variable. Its production term is dependent on KaiC which acts as a negative feedback on its production. As in \cite{teng_robust_2013}, we consider only active degradation and not dilution for mRNA as it is order of magnitude larger. Secondly, we add a production term for unphosphorylated KaiC from its mRNA. Finally, all forms of KaiC are assumed to have the same linear dilution rate and active degradation rate.

As in \cite{teng_robust_2013}, we consider on top of the wild type (WT) cyanobacterial circadian clock described above, the case of a mutant cyanobacteria lacking the transcriptional feedback. This is simply done by replacing the production term of mRNA by a constant production rate $V_{sptr}$. This mutant is referred to as post-transcriptional (PTR) mutant while the WT clock has a transcriptional translational regulation (TTR).

\begin{figure*}[ht!]
\includegraphics[width = \textwidth]{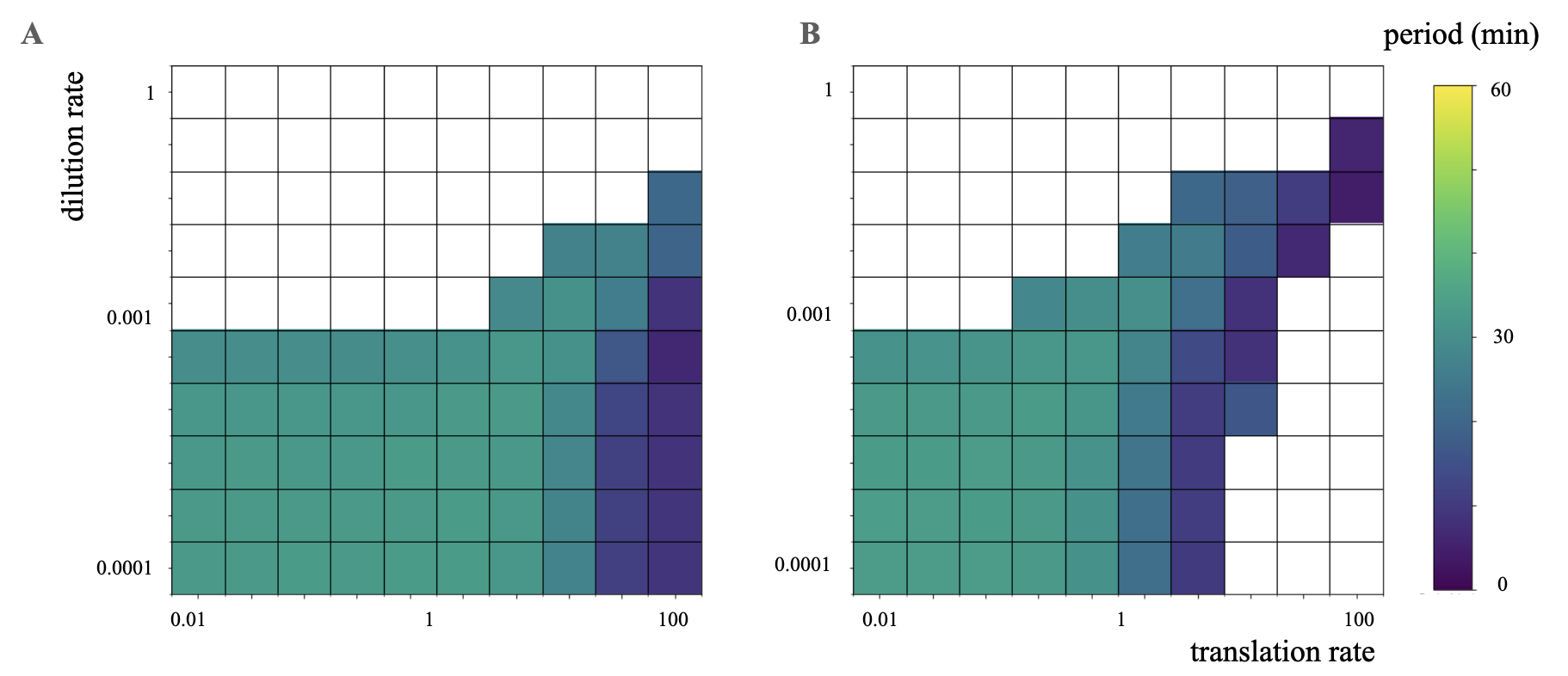}
\caption{\label{fig:ptr_vs_wt} Robustness analysis of the wild type model (panel A) and the model for the mutant lacking the transcriptional regulation (panel B) over parameter changes. We varied both the translation rate $K_s$ and the dilution rate $V_d$ over 4 orders of magnitudes. }
\end{figure*}

We analyzed the sensitivity to parameter changes of the PTR mutant and showed that model does not requires fine tuning to generated oscillations, see Fig. \ref{fig:ptr_vs_wt}. We observe that the main change between the WT cyanobacteria and the mutants lacking the TTR circuit is that the WT bacteria are more robust to changes in the translation rate. However, the TTR circuit is not crucial for the clock to be robust against changes in the dilution rate. 

\section{Discussion}

We showed that our revised  \textit{in vivo}  model of the cyanobacterial circadian clock is compatible with the robust oscillations of the cyanobacteria mutant lacking the TTR feedback. This result suggests that although the TTR feedback enhances the robustness of the clock, the effect is not as strong as initially suggested. Our conclusion is in agreement with the experiments performed with higher growth rates for (see Fig. S4 of \cite{teng_robust_2013}). Indeed, those experiments show that the mutants lacking the TTR regulation are able to generate oscillations in an extended range of growth rate. Although the growth rate of the bacteria is not explicitly modeled, it will impact the dilution rate. In the original model of the \textit{in vivo} clock, the dilution rate needed to be finely tuned. The TTR regulation could actually play a more important role if the nonlinearity of the regulator terms was increased \cite{ferrell_modeling_2011,gonze_goodwin_2013}. Coordination of both mechanisms is key to enhance the robustness too. 
To sum up, we propose a revisited version of the \textit{in vivo} cyanobacterial clock which is in agreement with current experiments. 
We expect that the synchronysation properties of the clock will also be enhanced by TTR regulation. 
In the future, we should to include the key proteins $RpaA$ and $SasA$ in the models, to be able to describe quantitatively how the clock's time is read as an output \cite{zwicker_robust_2010,tseng_structural_2017,chavan_reconstitution_2021}. \newline

Python codes to reproduce all results are available at \url{https://github.com/sophiedeb/cyano_clock_models}.


\section{Appendix : Reaction rates depend on the state of the system}

To obtain the dependence of the reaction rates on the various phosphoforms of KaiC, we need to describe the different allosteric states of KaiC. KaiC can be in a state comptent to bind KaiA and a state competent to bind KaiB, respectively denoted $\text{KaiC}^A$ and $\text{KaiC}^B$ (to simplify notations, we omit the $i,j,k$ indices those allosteric state should carry). From the state comptent to bind KaiA, KaiC can be converted in the active form of KaiC, which we denote KaiC${}^{*}_{i,j,k}$ with indices specifying the phosphoform. The allosteric reactions are considered to be at equilibrium, and one can therefore obtain the ratio of KaiC in its different allosteric states. The fact that those ratios are depend on the specific $(i,j,k)$-phosphoform is key to the \textit{in vitro} feedback mechanism. The (de)phosphorylation reaction rates are given by:
\begin{equation} 
k^{i,j,k}_{XY} = F^{i,j,k}_A \, k^A_{XY} \, + \, (1-F_A^{i,j,k} ) \, k^0_{XY} , \label{phosphorate}
\end{equation} 
where $F^{i,j,k}_A$ denotes the fraction of KaiC in the state competent to bind KaiA. 
The first term is dominant as KaiC autokinase activities are enhanced by KaiA. The constants $k^A_{XY}$ denotes the maximal phosphorylation rate from the state $X$ to the state $Y$ when KaiA  is bound, with $X,Y$ representing the $U,T,S$ or $D$ state on the relevant subunit. Similarly, $k^0_{XY}$ denotes the maximal phosphorylation rate when KaiA is not bound. The dependence of $F_A^{i,j,k}$ on the state of the system is given by
\begin{eqnarray}
 F_A^{i,j,k}   &\equiv& \frac{\text{KaiC}_{i,j,k}}{\text{KaiC}^B + \text{KaiC}^A + \text{KaiC}_{* i,j,k}} \nonumber  \\
 &\overset{\text{ \tiny equilibrium}}{=}& \frac{\text{KaiA}}{\text{KaiA} + K_m K_A^{i,j,k} + K_m} \nonumber
 \end{eqnarray}
where $K_m = \frac{k_2}{k_1} $ is the dissociation constant of the reaction KaiC${}^A \xrightleftharpoons[k_1 \text{KaiA}]{k_2} \text{KaiC}^{*}_{i,j,k}$, and $K_A^{i,j,k} = e^{- \frac{\Delta G_{i,j,k}}{k_B T}}$ the dissociation constant of the allosteric transformation KaiC${}^B \rightleftharpoons$ KaiC$^A$, with $\Delta G_{i,j,k} = i \, \Delta G_{pT} + j \, \Delta G_{pS} + k \, \Delta G_{pSpT} + (n-(i+j+k)) \Delta G_{U} $ being the free energy difference between the allosteric states. 

The (un)binding reactions of KaiB are dependent on 
the fraction of KaiC in form competent to bind to KaiB which is given by 
\begin{eqnarray} 
F_B^{i,j,k} &\equiv& \frac{ \text{KaiC}^B }{\text{KaiC}^B + \text{KaiC}^A + \text{KaiC}^{* i,j,k}} \label{bindrate}  \\
&=& \frac{1}{1 + \frac{1}{K_A^{i,j,k}} + \frac{\text{KaiA}}{K_m K_A^{i,j,k}}} \nonumber
\end{eqnarray}
Finally, we should note that KaiC$_{i,j,k}$ in Eqs.~\ref{Eq.invitro} represents the sum of all allosteric forms of KaiC in the $(i,j,k)$ state. 


\begin{table}[h]
\caption{ Parameter values - taken from \cite{rust_ordered_2007,teng_robust_2013,lin_mixtures_2014} \label{table:params}}
\begin{ruledtabular}
\begin{tabular}{llll}
$k_{on,B}$ & 0.15  $h^{-1}$ &
$k_{off,B}$ & $3.0  \,  10^{-2}$ $h^{-1}$\\
$\Delta G_{pT}$ &2.0 k${}_\text{B}$ T&
$\Delta G_{pS} $ &-3.5 k${}_\text{B}$ T\\
$ \Delta G_{pSpT}$ &-1.0 k${}_\text{B}$T&
$\Delta G_U$ &1.0  k${}_\text{B}$ T\\
$k_m$ &0.43  $\mu$ M  &
$kaiA0$ &10$\mu$ M \\
$k^0_{UT}$ &0.0   $h^{-1}$ &
$k^0_{TD}$ &0.0   $h^{-1}$\\
$k^0_{US}$ &0.0   $h^{-1}$ &
$k^0_{SD}$ &0.0   $h^{-1}$\\
$k^0_{DS}$ &0.31   $h^{-1}$&
$k^0_{DT}$ &0.0   $h^{-1}$ \\
$k^0_{TU}$ &0.21   $h^{-1}$ &
$k^0_{SU}$ &0.11    $h^{-1}$ \\
$k^{act}_{UT}$ &0.48   $h^{-1}$ &
$k^{act}_{TD}$ &0.21   $h^{-1}$\\
$k^{act}_{US}$ & $5.32 \, 10^{-2}$   $h^{-1}$ &
$k^{act}_{SD}$ &0.506   $h^{-1}$  \\
$k^{act}_{DS}$ &0.0   $h^{-1}$&
$k^{act}_{DT}$ &0.172   $h^{-1}$ \\
$k^{act}_{TU}$ &0.29   $h^{-1}$ &
$k^{act}_{SU}$ &0.9   $h^{-1}$ \\
$V_{sptr}$ & $3.316 \, 10^{-2}$ $\mu M h^{-1}$ &
$V_s$ &0.004 $\mu$ M  $h^{-1}$ \\
$V_m$ &0.20 $\mu$ M   $h^{-1}$  &
$K_m$ &0.20 $\mu$ M \\
$K$ &2.0 $\mu$ M  &
$V$ & $2.0 \, 10^{-3} $ $\mu$ M   $h^{-1}$\\
$K_i$ &1.0$\mu$ M&  
$K_s$ &varied \\
$V_d$ & varied 
\end{tabular}
\end{ruledtabular}
\end{table}

\section{acknowledgments}
SdB would like to warmly thank Susan Golden for an insightful discussion.


%

\end{document}